\documentstyle[12pt,epsfig,axodraw,a4]{article} 
\def\bild#1#2{    
        \vspace*{-5mm}
        \begin{center}
        \begin{math}
        \epsfxsize#2cm
        \epsffile{#1}
        \end{math}
        \end{center}  }
\newcommand{\vs}{\vspace{-0.25cm}}
\begin{document} 
\begin{center}
\large{\bf Chiral 3\begin{boldmath}$\pi$\end{boldmath}-exchange NN-potentials:
Results for dominant next-to-leading order contributions}

\medskip 

\bigskip

N. Kaiser\\

\bigskip

Physik Department T39, Technische Universit\"{a}t M\"{u}nchen,\\
    D-85747 Garching, Germany

\end{center}

\bigskip

\begin{abstract}
We calculate in (two-loop) chiral perturbation theory the local NN-potentials
generated by the three-pion exchange diagrams with one insertion from the
second order chiral effective pion-nucleon Lagrangian proportional to the
low-energy constants $c_{1,2,3,4}$. The resulting isoscalar central potential
vanishes identically. In most cases these $3\pi$-exchange potentials are larger
than the ones generated by the diagrams involving only leading order vertices 
due to the large values of $c_{3,4}$ (which mainly represent virtual 
$\Delta$-excitation). A similar feature has been observed for 
the chiral $2\pi$-exchange. We also give suitable (double-integral)
representations for the spin-spin and tensor potentials generated by the
leading-order diagrams proportional to $g_A^6$ involving four nucleon
propagators. In these cases the Cutkosky rule cannot be used to calculate the
spectral-functions in the infinite nucleon mass limit since the corresponding 
mass-spectra start with a non-vanishing value at the $3\pi$-threshold. 
Altogether, one finds that chiral $3\pi$-exchange leads to small corrections in
the region $r\geq 1.4$\,fm where $1\pi$- and chiral $2\pi$-exchange alone
provide a very good strong NN-force as shown in a recent analysis of the
low-energy pp-scattering data-base.       
\end{abstract}


\bigskip
To be published in {\it The Physical Review C (2000)}
\bigskip

Over the last years effective field theory methods have been successfully
applied to the two-nucleon system at low and intermediate energies 
\cite{weinb,kolck,kaplan,epelb1,epelb2,nnpap1,nnpap2}. The idea of constructing
the NN-potential from effective field theory was put forward by Weinberg
\cite{weinb} and this  was taken up by van Kolck and collaborators
\cite{kolck}. Their NN-potential contained one- and two-pion exchange graphs
based on time-ordered perturbation theory (regularized by a Gaussian cut-off)
and a host of contact interactions. The method of unitary transformations was
used by Epelbaoum, Gl\"ockle and Mei{\ss}ner \cite{epelb1} to construct an
energy-independent NN-potential from the effective chiral pion-nucleon
Lagrangian. Based on one- and two-pion exchange and a few contact interactions
which contribute only to S- and P-waves a good description of the deuteron
properties as well as the NN phase-shifts and mixing angles below
$T_{lab}=300$\,MeV was found in that framework \cite{epelb2}.   
 
The issue of the chiral $3\pi$-exchange was first addressed by Pupin and
Robilotta \cite{robil} who considered one particularly simple diagram with no
internal nucleon propagators. In a recent work \cite{3pipot} we have started 
to calculate the static NN-potentials generated by the (two-loop)
$3\pi$-exchange diagrams with all possible interaction vertices taken from the
leading-order effective chiral $\pi N$-Lagrangian. In ref.\cite{3pipot} we
restricted ourselves to the evaluation of four relatively simple classes of
diagrams with the common  prefactor $g_A^2/f_\pi^6$. It was stressed in
ref.\cite{3pipot} that the chiral $3\pi NN$-contact vertex employed in
ref.\cite{robil} depends on the choice of the interpolating pion-field  
and therefore one has to consider representation invariant classes of diagrams
by supplementing graphs involving the chiral $4\pi$-vertex. Another obvious 
consequence of this is that there is no unique definition of the so-called 
"pion-nucleon form factor" which is often introduced in phenomenological 
models. The only meaningful separation of the NN-potential is the one into 
point-like $1\pi$-exchange and $3\pi$-exchange. The four classes of 
$3\pi$-exchange diagrams evaluated in ref.\cite{3pipot} gave rise only to 
isovector spin-spin and tensor potentials with a strength of $\pm 0.1$\,MeV or
less at $r=1.0\,$fm. Compared to the chiral $2\pi$-exchange NN-potentials 
\cite{nnpap1,nnpap2} these are negligibly small corrections. 

In a subsequent work \cite{dreip} the static NN-potentials generated by the 
$3\pi$-exchange diagrams proportional to $g_A^4$ and $g_A^6$ (grouped into
classes\,V,\,VI,\,VII,\,VIII and IX) have been worked out. The resulting 
coordinate space potentials have typical strengths of a few MeV at $r=1.0$\,fm.
A surprising result of ref.\cite{dreip} was that the total isoscalar central 
$3\pi$-exchange NN-potential vanishes identically. The static $3\pi$-exchange 
NN-potentials have been constructed in refs.\cite{3pipot,dreip} from the 
mass-spectra (i.e. the imaginary parts) of the respective two-loop diagrams in
the infinite nucleon mass limit $M\to \infty$. This method could however not 
be successfully applied to the spin-spin and tensor potentials generated by
the  diagrams proportional to $g_A^6$ (classes\,VIII and IX) involving four
nucleon propagators. In these cases the integrand of the $3\pi$-phase space
integral $\int\int_{z^2\leq 1}d\omega_1 d\omega_2\dots$ developed  a
non-integrable singularity of the form $(1-z^2)^{-3/2}$ on the boundary
$z^2=1$. By considering the analytically solvable example of the crossed
three-boson exchange with scalar coupling in ref.\cite{dreip} the failure of
the spectral-function method could be understood from the fact that in the
limit $M\to \infty$ the mass-spectrum starts with a non-vanishing value at
the 3-boson threshold $\mu=3m$. In coordinate space this feature manifests
itself as a large-$r$ asymptotics of the form $e^{-3m r}r^{-2}$.    

The purpose of this work is to present suitable double-integral representations
for these remaining $3\pi$-exchange spin-spin and tensor NN-potentials
proportional to $g_A^6$ together with numerical results. Furthermore, will work
out here the next-to-leading order corrections to the chiral $3\pi$-exchange 
NN-potential. In general, the next-to-leading order contributions are given 
by all diagrams with one insertion from the second order chiral 
$\pi N$-Lagrangian ${\cal L}_{\pi N}^{(2)}$. However, from an analogous 
calculation of the chiral $2\pi$-exchange in ref.\cite{nnpap1} one has 
experienced that the dominant contributions come from the second order chiral 
$\pi\pi NN$-contact vertex proportional to the low-energy constants 
$c_{1,2,3,4}$, whereas the relativistic $1/M$-corrections are much smaller. 
The exception to this is, of course, the spin-orbit NN-potential which is a 
truly relativistic effect vanishing in the infinite nucleon mass limit $M\to
\infty$. Therefore we will consider here only the two-loop $3\pi$-exchange
diagrams with exactly one insertion proportional to $c_{1,2,3,4}$. The
corresponding  effective chiral $\pi N$-Lagrangian reads
\begin{equation} {\cal L}_{\pi N}^{(2)} = \bar N \Big\{ 2c_1 \,m_\pi^2
(U+U^\dagger) +c_2\, u_0^2+c_3\, u_\nu u^\nu +{i\over 2}\,c_4\,\vec \sigma\cdot
(\vec u \times \vec u\,) \Big\} N \,,\end{equation}
with the axial vector quantity $u^\nu = i \{\xi^\dagger, \partial^\nu \xi\}$.
The unitary matrix $\xi^2=U= 1+i\,\vec \tau\cdot \vec \pi/f_\pi-\vec 
\pi\,^2/2f_\pi^2 +\dots $ collects the Goldstone pion-fields and
$f_\pi=92.4$\,MeV denotes the weak pion decay constant. The low-energy
constants $c_{1,2,3,4}$ have been determined in a one-loop chiral perturbation
theory analysis of low-energy $\pi N$-scattering in the physical region 
\cite{nadja} and inside the Mandelstam triangle \cite{buettik}. Their
central values are $c_1=-0.8$\,GeV$^{-1}$, $c_2=3.2$\,GeV$^{-1}$, 
$c_3=-4.7$\,GeV$^{-1}$ and $c_4=3.4$\,GeV$^{-1}$. We prefer to use the values
obtained in ref.\cite{buettik} since at the small sub-threshold pion-energies 
relevant inside the Mandelstam triangle the chiral expansion of the $\pi
N$-scattering amplitude is expected to work best. As outlined 
in ref.\cite{aspects} the large low-energy constants $c_{2,3,4}$ receive 
dominant contributions from virtual $\Delta(1232)$-isobar excitation. In 
ref.\cite{nnpap2} it has been demonstrated that the $2\pi$-exchange 
NN-potentials generated by explicit $\Delta(1232)$-isobar excitation are very 
similar to those obtained  from the $NN\pi\pi$-contact interaction 
\cite{nnpap1} proportional to $c_{1,2,3,4}$. In the so-called small scale
expansion of ref.\cite{kamb} the $\Delta N$ mass-splitting $\Delta =293$\,MeV
is treated as a further small expansion parameter and $2\pi$-exchange graphs
with $\Delta$-excitation are counted as leading order ones. This is another
reason to focus on the $3\pi$-exchange potential proportional to $c_{1,2,3,4}$.
The second order chiral $\pi\pi NN$-contact vertex following from eq.(1) reads
\begin{equation} {i \over f_\pi^2}\Big\{  -2\delta^{ab} \big[ 2c_1 \,m_\pi^2 
+c_2\,q_a^0 q_b^0 +c_3\, q_a\cdot q_b\big] + c_4 \, \epsilon^{abc} \tau^c \,
\vec \sigma \cdot (\vec q_a\times \vec q_b\, ) \Big\}\,,  \end{equation}
where both pion four-momenta $q_a$ and $q_b$ are out-going ones. Note that the
isoscalar part in eq.(2) is spin-independent while the isovector part is 
spin-dependent.   

Next, let us give some basic definitions in order to fix our notation. In the
static limit and considering only irreducible diagrams the on-shell NN T-matrix
takes the following form
\begin{eqnarray} {\cal T}_{NN} &=& V_C(q) + V_S(q)\,\vec \sigma_1\cdot \vec 
\sigma_2 + V_T(q) \,\vec \sigma_1\cdot \vec q \,\,  \vec \sigma_2 \cdot \vec q
\nonumber \\ && + \Big[ W_C(q) + W_S(q)\,\vec \sigma_1\cdot  \vec \sigma_2
+ W_T(q) \,\vec \sigma_1\cdot \vec q \,\,  \vec \sigma_2 \cdot \vec q \, \Big] 
\,\vec \tau_1 \cdot \vec \tau_2\,, \end{eqnarray} 
where $q=|\vec q\,|$ denotes the momentum transfer between the initial and
final-state nucleon. The subscripts $C,S$ and $T$ refer to the central, 
spin-spin and tensor components, each of which occurs in an isoscalar
($V_{C,S,T}$) and an isovector version ($W_{C,S,T}$). As indicated, the (real) 
NN-amplitudes $V_C(q),\dots, W_T(q)$ depend only on the momentum transfer $q$ 
in the static limit $M\to \infty$. We are here interested only in the 
coordinate-space potentials generated by certain diagrams in which three pions 
are simultaneously exchanged between both nucleons. For this purpose it is 
sufficient to calculate the imaginary parts of the NN-amplitudes,
Im\,$V_{C,S,T} (i\mu)$ and Im\,$W_{C,S,T}(i\mu)$, analytically continued to  
time-like momentum transfer $q=i\mu-0^+$ with $\mu\geq 3m_\pi$. These imaginary
parts are then the mass-spectra entering a representation of the local 
coordinate-space potentials in the form of a continuous superposition of 
Yukawa-functions,     
\begin{eqnarray} \widetilde V_C(r) &=& -{1\over 2\pi^2 r} \int_{3m_\pi}^\infty
d\mu \,\mu \,e^{-\mu r} \, {\rm Im}\, V_C(i\mu)\,, \\ \widetilde V_S(r) &=& 
{1\over 6\pi^2 r} \int_{3m_\pi}^\infty d\mu \,\mu \,e^{-\mu r} \Big[ \mu^2\,
{\rm Im}\, V_T(i\mu) - 3\, {\rm Im}\,V_S(i\mu) \Big]\,, \\ \widetilde V_T(r)
&=& {1\over 6\pi^2 r^3} \int_{3m_\pi}^\infty d\mu\,\mu\, e^{-\mu r}(3+3\mu r+
\mu^2r^2){\rm Im}\,V_T(i\mu)\,.\end{eqnarray}
The isoscalar central, spin-spin and tensor potentials, denoted here by
$\widetilde V_{C,S,T}(r)$, are as usual those ones which are accompanied by the
operators $1$, $\vec \sigma_1\cdot \vec \sigma_2$ and  $3\,\vec \sigma_1\cdot
\hat r\,\vec \sigma_2 \cdot \hat r -\vec \sigma_1\cdot \vec \sigma_2$,
respectively. For the isovector potentials $\widetilde W_{C,S,T}(r)$ a
completely analogous representation holds. The imaginary parts entering
eqs.(4,5,6) are calculated from the pertinent 2-loop $3\pi$-exchange diagrams 
as integrals of the $\bar NN\to3\pi \to \bar NN$ transition amplitudes over 
the Lorentz-invariant $3\pi$-phase space making use of the Cutkosky
cutting rule as explained in refs.\cite{3pipot,dreip,spectral}. The sub-leading
$3\pi$-exchange considered here occurs at fifth order in small momentum (or 
chiral) expansion. Of course at this order there are also many two-loop 
$2\pi$-exchange diagrams proportional to $c_{1,2,3,4}$. Since the 
corresponding spectral-functions will (in general) contain one-loop
sub-divergences (related to the fourth order chiral $\pi N$-amplitude) an 
adequate discussion of these $2\pi$-exchange potentials requires good knowledge
of the strength of the counter-terms $\bar e_i$ in ${\cal L}^{(4)}_{\pi N}$
\cite{nadja2}. With these included the corresponding $2\pi$-exchange potential
is well-defined, however, its explicit computation goes far beyond the scope of
the present paper.  

We start with the diagrams of class\,X shown in Fig.1. The heavy dot symbolizes
the second order chiral $\pi\pi NN$-contact vertex eq.(2) proportional to
$c_{1,2,3,4}$. Diagrams for which the role of both nucleons is interchanged are
not shown. These lead to the same contribution to the NN-potential, i.e. a
factor of 2. As stressed in ref.\cite{3pipot} diagrams involving the chiral 
$3\pi  NN$-vertex or the chiral $4\pi$-vertex depend on an arbitrary parameter
and therefore one should consider the full class\,X as one entity. Obviously,
the  last two pion-pole diagrams in Fig.1 contribute via coupling constant 
renormalization also to the point-like $1\pi$-exchange. This effect is however 
automatically taken care by working with the physical (on-shell) 
$\pi NN$-coupling constant $g_{\pi N}$. 

\bigskip
\medskip

\bild{figu10.epsi}{16}
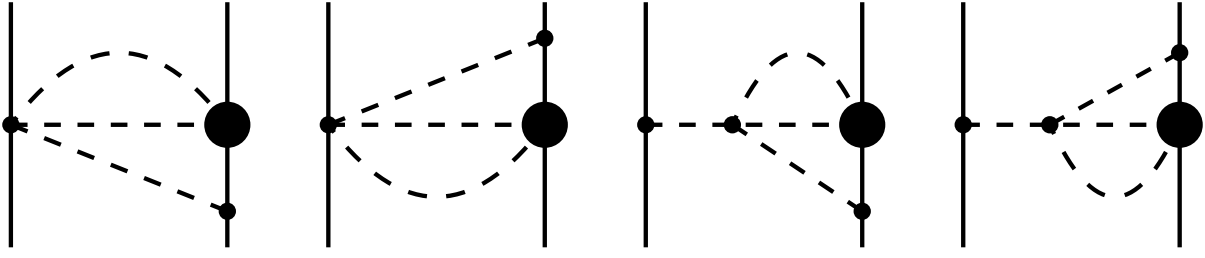
{\it Fig.1: $3\pi$-exchange diagrams of class\,X proportional to $g_A^2$. 
Solid and dashed lines represent nucleons and pions, respectively. The heavy
dot symbolized an insertion from second order chiral $\pi N$-Lagrangian. The
combinatoric factor of these diagrams is 1/2. Diagrams for which
the role of both nucleons is interchanged are not shown. They lead to the same
NN-potential.}

\bigskip

Note that we are considering here only 
the finite-range Yukawa-parts of the $1\pi$-exchange and we disregard all 
zero-range $\delta^3(\vec r\,)$-terms. Transformed into momentum space the 
latter become polynomials in $q^2$ with possible contributions from
higher-derivative operators. From an inspection of the spin- and isospin 
factors occurring in the diagrams of class\,X one finds immediately that only 
non-vanishing isovector spin-spin and tensor NN-amplitudes $W_{S,T}$ will be 
obtained. We find the following imaginary parts from class\,X (dropping from
now on the argument $i\mu$), 
\begin{eqnarray} {\rm Im}\, W_S^{(X)} &=& {g_A^2 \over (4f_\pi)^6 \pi^2
\mu^3} \int_{2m_\pi}^{\mu-m_\pi} \!\!dw \sqrt{w^2-4m_\pi^2}\, \Big\{ \Big[ 
2C_{13}(w)+{c_2\over 3}(w^2-4m_\pi^2)\Big] \lambda(w) \nonumber \\ & &  
+{c_4\over3}(w^2-4m_\pi^2)\Big[(w^2-m_\pi^2)^2
+\mu^2 (2m_\pi^2+2w^2-3\mu^2) \Big] \Big\} \,,\end{eqnarray}    
\begin{eqnarray} {\rm Im}\, W_T^{(X)} &=&  {1\over \mu^2}\, {\rm Im}\, 
W_S^{(X)}+{g_A^2 \over (4f_\pi)^6 \pi^2 \mu^5} \int_{2m_\pi}^{\mu-m_\pi}
\!\!dw \sqrt{w^2-4m_\pi^2}\, \Big\{ \Big[4C_{13}(w)+{2c_2\over 3}(w^2-4m_\pi^2)
\Big] \nonumber \\ & & \times\,(\mu^2+m_\pi^2-w^2)\Big[\mu^2+m_\pi^2-w^2+2\mu^2
(2w^2-\mu^2)(\mu^2-m_\pi^2)^{-1} \Big] \nonumber \\ & &+{2c_4 \over3}(w^2-
4m_\pi^2)\lambda(w)(m_\pi^2+3\mu^2)(m_\pi^2-\mu^2)^{-1} \Big\}\,,\end{eqnarray}
where we have introduced the abbreviations $C_{13}(w)=4c_1m_\pi^2+c_3(w^2-2
m_\pi^2)$ and $\lambda(w) = w^4+\mu^4+m_\pi^4-2 w^2 \mu^2-2w^2 m_\pi^2-2 \mu^2 
m_\pi^2$. The variable $w$ denotes the invariant mass of a pion-pair and its
kinematically allowed range is $2m_\pi\leq w\leq \mu-m_\pi$. The $dw$-integrals
in eqs.(7,8) could of course be solved easily in terms of square-root and 
logarithmic functions. However, we want to avoid the resulting rather lengthy 
expressions. 

\bigskip
\bigskip

\bild{figu11.epsi}{16}
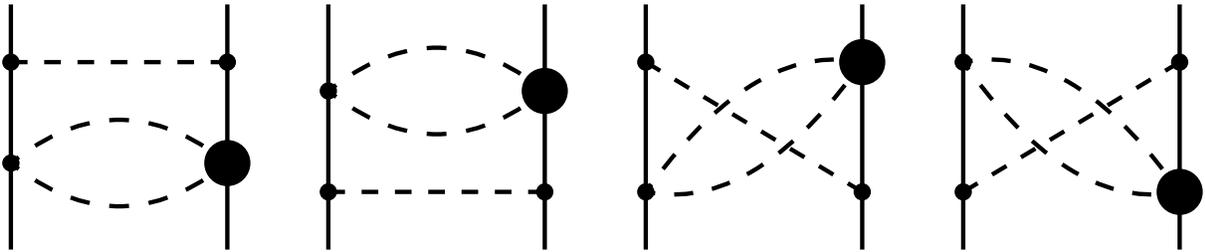
{\it Fig.2: $3\pi$-exchange diagrams of class\,XI proportional to $g_A^2$. 
These give a vanishing contribution to the NN-potential.}

\bigskip

Next, we consider the diagrams of class\,XI shown in Fig.2. Each of them gives
a vanishing contribution to the NN T-matrix for the following reason. Since the
leading order (Weinberg-Tomozawa) $\pi\pi NN$-vertex is of isovector nature
(i.e. proportional to $\epsilon^{abc}\tau^c$) a non-zero isospin-factor is
obtained only from the spin-dependent isovector term in eq.(2) proportional to 
$c_4$. When combined with the spin-independent Weinberg-Tomozawa vertex on the
other side of the bubble-type sub-diagram one finds immediately that the 
pertinent 1-loop integral is zero in the heavy baryon formalism (basically
because the spin-vector $\vec \sigma$ has no time-component). 

\bigskip
\bigskip

\bild{figu12.epsi}{16}
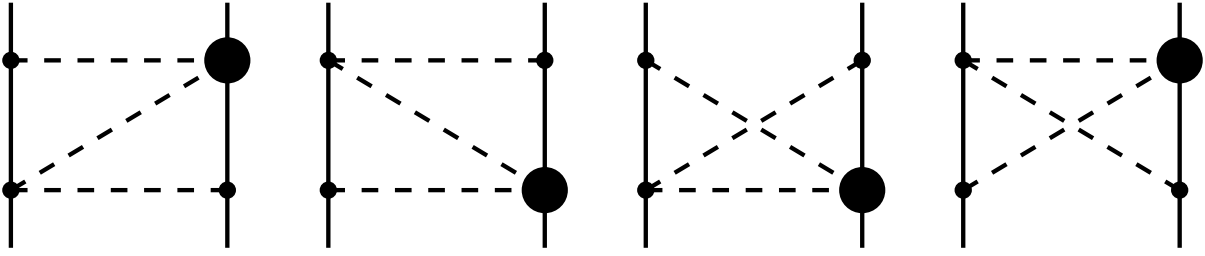
{\it Fig.3: $3\pi$-exchange diagrams of class\,XII proportional to $g_A^2$. For
further notation, see Fig.\,1.}

\bigskip

Next, we consider the diagrams of class\,XII shown in Fig.3. The isoscalar 
contribution comes exclusively from the $c_4$-term in eq.(2). Altogether one
obtains the following imaginary parts of the isoscalar and isovector spin-spin 
and tensor NN-amplitudes from class\,XII,     
\begin{equation} {\rm Im}\, V_S^{(XII)} = {2g_A^2c_4 \over (4f_\pi)^6 
\pi^2 \mu^3} \int_{2m_\pi}^{\mu-m_\pi} \!\!dw \sqrt{w^2-4m_\pi^2}\, \lambda(w)
\Big[7m_\pi^2-\mu^2-3 w^2+2m_\pi^2(m_\pi^2-\mu^2)w^{-2}\Big]\,,\end{equation}
\begin{eqnarray} {\rm Im}\, V_T^{(XII)} &=&  {1\over \mu^2}\, {\rm Im}\, 
V_S^{(XII)} +{4g_A^2 c_4\over (4f_\pi)^6 \pi^2 \mu^5} \int_{2m_\pi}^{\mu-
m_\pi} \!\!dw\sqrt{w^2-4m_\pi^2}\, (\mu^2+m_\pi^2-w^2) \nonumber \\ & & \times 
\Big[3w^4-2w^2(5m_\pi^2+2\mu^2)+\mu^4+2\mu^2m_\pi^2 +5m_\pi^4+2m_\pi^2
(\mu^2-m_\pi^2)^2 w^{-2} \Big] \,,\end{eqnarray}
\begin{eqnarray} {\rm Im}\, W_S^{(XII)} &=& {g_A^2 \over (4f_\pi)^6 \pi^2
\mu^3} \int_{2m_\pi}^{\mu-m_\pi} \!\!dw \sqrt{w^2-4m_\pi^2}\, \Big\{ \Big[
2C_{13}(w) +{c_2\over 3}(w^2-4m_\pi^2) \Big]\lambda(w) \nonumber \\ & &
+{c_4\over3}(w^2-4m_\pi^2)\Big[(\mu^2-m_\pi^2)
(3\mu^2+m_\pi^2-2w^2) -w^4\Big] \Big\}\,, \end{eqnarray}    
\begin{eqnarray} {\rm Im}\, W_T^{(XII)} &=&  {1\over \mu^2}\, {\rm Im}\, 
W_S^{(XII)} +{g_A^2 \over (4f_\pi)^6\pi^2\mu^5} \int_{2m_\pi}^{\mu-m_\pi}
\!\!dw\sqrt{w^2-4m_\pi^2}\,\Big\{ [(w^2-m_\pi^2)^2-\mu^4] \nonumber \\ & & 
\times\, \Big[4C_{13}(w) +{2c_2\over 3}(w^2-4m_\pi^2) \Big] -{2c_4 \over3}
(w^2-4m_\pi^2)\lambda(w) \Big\} \,.\end{eqnarray}

\bigskip
\bigskip

\bild{figu13.epsi}{16}
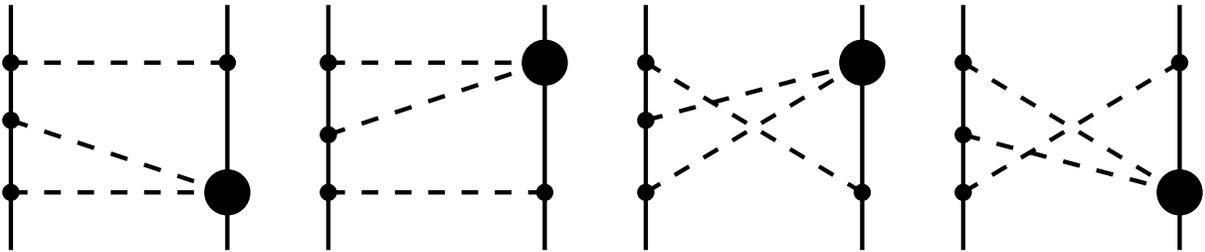
{\it Fig.4: $3\pi$-exchange diagrams of class\, XIII proportional to $g_A^4$.
For further notation, see Fig.\,1.}

\bigskip

Next, we turn to the diagrams of class\,XIII shown in Fig.4. The technique to
separate off the iterative part from the first two reducible diagrams has been
explained in ref.\cite{dreip}. The iterative (or reducible) part is defined
here (entirely) within perturbation theory as that part which carries in the
numerator the large scale enhancement factor $M$, the nucleon mass. Such an
iterative part does therefore not obey the naive chiral power counting rules 
(see also section 4.3 in ref.\cite{nnpap1} on the so-called iterated 
$1\pi$-exchange). A detailed discussion of possible ambiguities showing up in
(non-perturbative) iterations to infinite orders (e.g. via a  Schr\"odinger
equation) can be found in ref.\cite{friar}. The isovectorial spin-dependent 
contact vertex proportional to $c_4$ (and in fact only this one) produces now
also a central NN-amplitude. Interestingly, its isoscalar and isovector
components come with a fixed ratio. The corresponding imaginary parts  read
\begin{equation} {\rm Im}\, V_C^{(XIII)} = -{3\over 4}\, {\rm Im}\,W_C^{(
XIII)}= {12g_A^4c_4 \over (4f_\pi)^6 \pi^2 \mu} \int_{2m_\pi}^{\mu-m_\pi}
\!\!dw  \sqrt{w^2-4m_\pi^2}\,\lambda(w)\,. \end{equation}
Furthermore, the isoscalar and isovector spin-spin and tensor NN-amplitudes 
generated by the diagrams of class\,XIII have the following imaginary parts,  
\begin{eqnarray} {\rm Im}\, V_S^{(XIII)} &=& {g_A^4c_4 \over (4f_\pi)^6 
\pi^2 \mu^3} \int_{2m_\pi}^{\mu-m_\pi} \!\!dw \sqrt{w^2-4m_\pi^2}\, \Big\{
7\mu^6-37 \mu^4 m_\pi^2+45 \mu^2 m_\pi^4 -15 m_\pi^6 +9w^6 \nonumber \\ & & 
-w^4(11\mu^2+37 m_\pi^2) +w^2(45m_\pi^4+32\mu^2 m_\pi^2 -5\mu^4)
+ 2m_\pi^2(\mu^2-m_\pi^2)^3w^{-2} \Big\}\,, \end{eqnarray}    
\begin{eqnarray} {\rm Im}\, V_T^{(XIII)} &=& {1\over \mu^2}\, {\rm Im}\, 
V_S^{(XIII)}+{2g_A^4 c_4\over (4f_\pi)^6 \pi^2 \mu^5} \int_{2m_\pi}^{\mu-
m_\pi} \!\!dw\sqrt{w^2-4m_\pi^2} \nonumber \\ & & \times\, \Big\{9w^6-w^4(
19\mu^2+37  m_\pi^2) +w^2(11\mu^4+52 \mu^2 m_\pi^2+45 m_\pi^4) \nonumber \\ & &
-\mu^6-13 \mu^4 m_\pi^2 -35 \mu^2m_\pi^4 -15m_\pi^6+2m_\pi^2
(m_\pi^2-\mu^2)(\mu^4-m_\pi^4) w^{-2} \Big\}\,,  \end{eqnarray}    
\begin{eqnarray} {\rm Im}\, W_S^{(XIII)} &=& {g_A^4 \over(4f_\pi)^6 \pi^2
\mu^3} \int_{2m_\pi}^{\mu-m_\pi} \!\!dw \sqrt{w^2-4m_\pi^2}\, \Big\{ 3C_{13}(w)
\Big[4\mu^2(\mu^2-m_\pi^2-w^2)-\lambda(w) \Big]  \nonumber \\ & &  
+{c_2\over2} (8m_\pi^2-5w^2)\lambda(w) + 4c_4 \Big[
w^6-w^4(\mu^2+4m_\pi^2) \nonumber \\ & & +w^2(\mu^4-2\mu^2 m_\pi^2+5m_\pi^4)
-\mu^6+4\mu^4 m_\pi^2 -\mu^2 m_\pi^4 -2m_\pi^6\Big] \Big\} \,,\end{eqnarray}   
\begin{eqnarray} {\rm Im}\, W_T^{(XIII)} &=&  {1\over \mu^2}\,{\rm Im}\, 
W_S^{(XIII)}+{g_A^4 \over (4f_\pi)^6\pi^2\mu^5} \int_{2m_\pi}^{\mu-m_\pi}
\!\!dw\sqrt{w^2-4m_\pi^2}\,\Big\{ (m_\pi^2+\mu^2 -w^2) \nonumber \\ & & \times
\Big[6C_{13}(w) (3\mu^2+w^2-m_\pi^2) +c_2(8m_\pi^2-5w^2)(m_\pi^2+\mu^2-w^2) 
\Big] \nonumber \\ & & +8c_4 (w^2-2m_\pi^2)[(w^2-m_\pi^2)^2-\mu^4] \Big\} \,.
\end{eqnarray} 

\bigskip
\bigskip

\bild{figu14.epsi}{8}
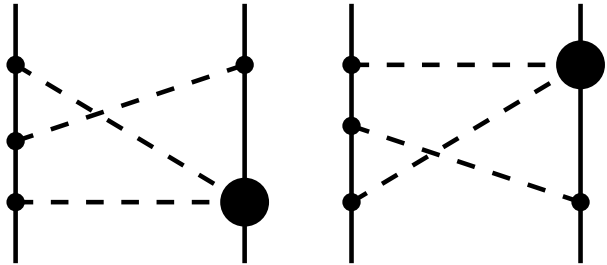
{\it Fig.5: $3\pi$-exchange diagrams of class\,XIV proportional to $g_A^4$. For
further notation, see Fig.\,1.} 

\bigskip

Finally, we consider the (irreducible) diagrams of class\,XIV shown in Fig.5. 
In this case the isovectorial $c_4$-term in eq.(2) does not make a contribution
to the isovector NN-amplitudes $W_{C,S,T}$. One obtains from class\,XIV an 
isoscalar  central NN-amplitude which is however exactly canceled by the one 
coming from  class\,XIII, 
\begin{equation}{\rm Im}\, V_C^{(XIV)}=-{\rm Im}\,V_C^{(XIII)}\,.\end{equation}
Therefore, one has the interesting result that there is no isoscalar central 
NN-potential from chiral $3\pi$-exchange even at next-to-leading order
(considering the dominant $c_{1,2,3,4}$-terms).  For the imaginary parts of 
the spin-spin and tensor NN-amplitudes generated by the diagrams of class\,XIV
one obtains the following expressions, 
\begin{eqnarray} {\rm Im}\, V_S^{(XIV)} &=& {g_A^4c_4 \over (4f_\pi)^6 
\pi^2 \mu^3} \int_{2m_\pi}^{\mu-m_\pi} \!\!dw \sqrt{w^2-4m_\pi^2}\, \Big\{
9w^6 -5\mu^6-13 \mu^4 m_\pi^2+33\mu^2 m_\pi^4 -15 m_\pi^6  \nonumber \\ & & 
-w^4(23\mu^2+37 m_\pi^2) +w^2(45m_\pi^4+8\mu^2 m_\pi^2 +19\mu^4)
+ 2m_\pi^2(\mu^2-m_\pi^2)^3w^{-2} \Big\} \,,\end{eqnarray}    
\begin{eqnarray} {\rm Im}\, V_T^{(XIV)} &=&  {1\over \mu^2}\, {\rm Im}\, 
V_S^{(XIV)} +{2g_A^4 c_4\over (4f_\pi)^6 \pi^2 \mu^5} \int_{2m_\pi}^{\mu-
m_\pi} \!\!dw\sqrt{w^2-4m_\pi^2} \nonumber \\ & & \times\, \Big\{9w^6-w^4(
19\mu^2+37  m_\pi^2) +w^2(11\mu^4+52 \mu^2 m_\pi^2+45 m_\pi^4) \nonumber \\ & &
-\mu^6-13 \mu^4 m_\pi^2 -35 \mu^2m_\pi^4 -15m_\pi^6+2m_\pi^2
(m_\pi^2-\mu^2)(\mu^4-m_\pi^4) w^{-2} \Big\}\,,  \end{eqnarray}    
\begin{eqnarray} {\rm Im}\, W_S^{(XIV)} &=& {g_A^4 \over (4f_\pi)^6 \pi^2
\mu^3} \int_{2m_\pi}^{\mu-m_\pi} \!\!dw \sqrt{w^2-4m_\pi^2}\, \lambda(w)  
\nonumber \\ & & \times \Big\{ -C_{13}(w) +{c_2\over6} \Big[ w^2+
2\mu^2-6 m_\pi^2+4m_\pi^2(\mu^2-m_\pi^2) w^{-2}\Big] \Big\} \,,\end{eqnarray}  
\begin{eqnarray} {\rm Im}\, W_T^{(XIV)} &=&  {1\over \mu^2}\, {\rm Im}\, 
W_S^{(XIII)}+{g_A^4 \over (4f_\pi)^6\pi^2\mu^5} \int_{2m_\pi}^{\mu-m_\pi}
\!\!dw\sqrt{w^2-4m_\pi^2}\,\Big\{ -2C_{13}(w)  \nonumber \\ & & \times 
(m_\pi^2+\mu^2 -w^2)^2 +{c_2\over 3}\Big[ w^6-4w^4(\mu^2+2m_\pi^2)+2(m_\pi^2-
\mu^2)^3  \nonumber \\ & &  +w^2(5 \mu^4 +6\mu^2 m_\pi^2+9m_\pi^2)+ 4 m_\pi^2
(\mu^4-m_\pi^4)(m_\pi^2-\mu^2)w^{-2} \Big] \Big\}\,. \end{eqnarray} 
This completes the presentation of analytical results. In Table\,1, we present 
numerical results for the coordinate-space NN-potentials generated by the 
$3\pi$-exchange graphs of classes\,X,\,XII,\,XIII and XIV for inter-nucleon 
distances 0.8\,fm$\,\leq r\leq \,$1.4\,fm. We use the parameters $f_\pi=92.4
$\,MeV, $m_\pi=138\,$MeV (average pion mass) and the central value of the 
nucleon axial-vector coupling-constant $g_A=1.267\pm 0.004$ \cite{pdg}. For the
second order low-energy constants $c_{1,2,3,4}$ we use the values $c_1=-0.8
$\,GeV$^{-1}$, $c_2=3.2$\,GeV$^{-1}$, $c_3=-4.7$\,GeV$^{-1}$, 
$c_4=3.4$\,GeV$^{-1}$ mentioned above. One observes from Table\,1 that the
next-to-leading  order $3\pi$-exchange NN-potentials are often larger than the
leading order ones (comparing e.g. values at $r=1.0$\,fm). The reason for this
are the large values of the low-energy constants $c_3$ and $c_4$ representing
the strong $\Delta$-excitation effects. A similar feature has also been 
observed for the chiral $2\pi$-exchange NN-potential in ref.\cite{nnpap1}. 
In that case the effects occured however more selectively in the various 
channels. For example, there was no $2\pi$-exchange isoscalar central potential
at leading order and at next-to-leading a large isoscalar central potential 
with a strength of about $-300$\,MeV at $r=1.0$\,fm was generated mainly by the
$\pi\pi NN$-contact vertex proportional to $c_3$. Note that the low-energy 
constant $c_3$ is related to the nucleon axial polarizability 
\cite{nnpap1,aspects}. The repulsive isovector central potential $\widetilde 
W_C^{(XIII)}(r)$ from class\,XIII with a strength of almost $50$\,MeV at $r=1.0
$\,fm is also fairly large. The largest potential is however the attractive 
isovector tensor potential $\widetilde W_T^{(XIII)}(r)$. At distances $r\leq 
1.1$\,fm it actually overwhelms the repulsive isovector tensor potential of the
(point-like) $1\pi$-exchange. (Since the eigen-values of $\vec \tau_1\cdot \vec
\tau_2$ are $-3$ and $+1$ the $1\pi$-exchange tensor potential can also act
attractively, e.g. in the $^3S_1$-state relevant for deuteron binding.) 
It is well-known that phenomenology requires a 
reduction of the $1\pi$-exchange NN-tensor force at intermediate distances. The
effect obtained here is presumably too strong and in addition the isovector 
tensor potential gets reduced by the chiral $2\pi$-exchange (see 
ref.\cite{nnpap1}). Because of the finite size (quark-substructure) of the 
nucleon one can trust the present calculation based on point-like chiral $\pi 
N$-interactions at most for distances $r\geq r_0 \simeq 1$\,fm. Clearly, as 
soon as the nucleons start to overlap substantially the picture of an
undisturbed multi-pion exchange between nucleons breaks down. The limiting 
distance scale $r_0\simeq 4/m_\rho \simeq 1$\,fm (with $m_\rho=770$\,MeV the
$\rho$-meson mass) should be considered as a  conservative estimate based on
the argument that the nucleon axial radius $r_A\simeq \sqrt{6}/m_\rho$ 
receives no substantial contributions from chiral pion-loops \cite{hemm}. Only
by confronting chiral multi-pion exchange potentials with NN-scattering data 
one is able to determine this limiting distance scale $r_0$ more precisely.

By analyzing the elastic proton-proton 
scattering data-base below $350$\,MeV laboratory kinetic energy it has been 
shown recently in ref.\cite{nijmeg} that $1\pi$- plus chiral
$2\pi$-exchange gives a very good strong NN-force at least as far inwards as 
$r=1.4$\,fm (i.e. the length scale corresponding to the pion Compton 
wave-length $1/m_\pi$). For such distances
($r\geq1.4$\,fm) the chiral $3\pi$-exchange NN-potential calculated here and 
previously in refs.\cite{3pipot,dreip}) turns out to be a negligibly small 
correction. By inspection of Table\,1 one observes that $3\pi$-exchange
potentials typically decrease by about a factor of 10 each $0.3$\,fm step
outwards.  

\bigskip
\bigskip

\bild{3pipotfig8.epsi}{8}
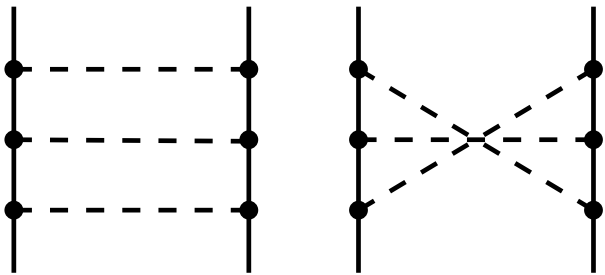
{\it Fig.6: $3\pi$-exchange diagrams of class\,VIII proportional to $g_A^6$.} 

\bigskip

Finally, we like to present results for the leading order spin-spin and tensor
NN-potentials generated by the $3\pi$-exchange diagrams proportional to $g_A^6$
(classes\, VIII and IX shown in Figs.6 and 7) which could not be evaluated with
the help of the spectral-function method in ref.\cite{dreip}. The spin- and 
isospin factors of the two diagrams of class\,VIII (shown in Fig.6) are such 
that the irreducible part of the first graph and the second graph give rise
only to a non-vanishing isoscalar central and isovector spin-spin and tensor
potential. For the latter the spectral-function method lead to non-integrable 
boundary singularities in the $3\pi$-phase space integral. Therefore an 
alternative calculational technique is required. 

The coordinate-space potential is generally 
given as the 3-dimensional Fourier transform $\int d^3q \exp(i\,\vec q \cdot
\vec r\,)\dots$ of the 2-loop momentum-space amplitude constructed from the 
propagators and vertices of a diagram. This requires in total 11 integrations,
$\int d^3q\,d^4l_1\,d^4l_2$. Usually, one performs first the integrations over
the loop-energies $l_{01}$ and $l_{02}$ using residue calculus and one obtains
this way the sum of energy denominators of all time-orderings of the
diagram. In the present case this would lead to a complicated rational function
of the three (on-shell) pion-energies $\omega_{1,2,3}$ which could not be
further integrated analytically. The strategy is to perform a Wick-rotation of
both loop-energies, $l_{0j}\to i\, l_{4j},\,j\in\{1,2\}$, (of course one has to
proof that this is indeed allowed) and to evaluate the remaining three
3-dimensional integrations analytically. The latter is possible because the
original integrand factorizes in the variables $\vec l_1$, $\vec l_2$ and $\vec
l_3= \vec q-\vec l_1+\vec l_2$ (see also the appendix in ref.\cite{dreip}). The
result of this involves among other factors (originating from the vertices of 
the  diagram) a product of three Yukawa-functions $\exp(-r \sqrt{m_\pi^2+
l^2_{4j}})/4\pi r$. This way one derives the following double-integral
representations for the isovector spin-spin and tensor potentials from 
class\,VIII,  
\begin{eqnarray} \widetilde W_S^{(VIII)}(r)&=& {7g_A^6 \over 6(16\pi f_\pi^2)^3
r^7} \bigg\{ e^{-3x}(8+18x+16x^2+6x^3+2x^4-x^5)\nonumber \\ &&- -\!\!\!\!\!\! 
\int_{-\infty}^{+\infty} {d\zeta_1d\zeta_2 \over \pi^2 \zeta_1\zeta_2 } \,
{\partial^2 \over \partial \zeta_1 \partial \zeta_2} \Big[e^{-x \Sigma} \Big(
12x^{-2}+12x^{-1}\Sigma+2\Sigma^2+8Q \nonumber \\ && +x
(2\Sigma Q+6\Pi)+2x^2 \Sigma \Pi-x^4 \Pi^2 \Big) \Big] \bigg\}\,,\end{eqnarray}
\begin{eqnarray} \widetilde W_T^{(VIII)}(r)&=& {7g_A^6 \over 6(16\pi f_\pi^2)^3
r^7} \bigg\{-e^{-3x}(1+x)(16+26x+18x^2+6x^3+x^4) \nonumber \\ && 
+ -\!\!\!\!\!\! \int_{-\infty}^{+\infty} {d\zeta_1d\zeta_2\over \pi^2 \zeta_1
\zeta_2 } \,{\partial^2 \over \partial \zeta_1 \partial \zeta_2} \Big[ e^{-x 
\Sigma}\Big(36x^{-2}+36x^{-1}\Sigma+10\Sigma^2+16Q \nonumber
\\ & & +x(10\Sigma Q+6\Pi)+x^2 (4\Sigma \Pi+3Q^2)+3x^3 Q \Pi+x^4 \Pi^2 \Big)
\Big] \bigg\} \,,\end{eqnarray}
where we have used the dimensionless variable $x = m_\pi r$ together with
various abbreviations: 
\begin{equation} \alpha =\sqrt{1+\zeta_1^2}\,,\qquad \beta =\sqrt{1+\zeta_2^2}
\,,\qquad\gamma = \sqrt{1+(\zeta_1-\zeta_2)^2}\,,\nonumber \\ \end{equation}
\begin{equation} \Sigma = \alpha+\beta+\gamma\,, \qquad Q = \alpha\beta+
\beta\gamma +\gamma \alpha\,, \qquad \Pi = \alpha\beta\gamma\,.\end{equation}
The asymptotic behavior $e^{-3m_\pi r}r^{-2}$ of $\widetilde W_{S,T}^{(VIII)}
(r)$ in eqs.(23,24) indicates that the corresponding mass-spectra start with a
non-vanishing value at the $3\pi$-threshold $\mu=3m_\pi$. For the numerical 
evaluation of the principal-value integrals in eqs.(23,24) it is advantageous
to employ the identity $-\!\!\!\!\!\int_{-\infty}^{+\infty}d\zeta \,f(\zeta)
\zeta^{-1} =\int_{0}^{+\infty} d\zeta \,[f(\zeta)-f(-\zeta)]\zeta^{-1}$, where
the right hand side is singularity-free. As a check we have derived an
analogous double-integral representation for the isoscalar central potential
$\widetilde V_C^{(VIII)}(r)$ and we have found perfect agreement with the 
numerical results of ref.\cite{dreip} based on the simpler spectral-function
method.

Lastly, we come to the diagrams of class\,IX shown in Fig.7. For the isoscalar
spin-spin and tensor NN-amplitudes the relevant sum of energy denominators is 
$(\omega_1^2+\omega_2^2)/(2\omega_1^4\omega_2^4\omega_3^2)$. Since its
denominator factorizes in the three variables $\vec l_1$, $\vec l_2$ and $\vec
l_3= \vec q-\vec l_1+\vec l_2$ one can perform the corresponding 9 integrations
analytically and one gets the following closed form expressions, 
\begin{equation} \widetilde V_S^{(IX)}(r)= {2g_A^6 \over (16\pi f_\pi^2)^3}
\,{e^{-3x}\over r^7} \,(2+2x+x^2)(2-2x-4x^2-x^3)\,, \end{equation}
\begin{equation} \widetilde V_T^{(IX)}(r)= {2g_A^6 \over (16\pi f_\pi^2)^3}
\,{e^{-3x}\over r^7}\,(2+2x+x^2)(2+x-x^2-x^3) \,.\end{equation}
We have corrected here in eq.(28) a small misprint (factor 2) which has occured
in eq.(24) of ref.\cite{dreip}. The same features concerning factorization hold
for the isovector central NN-amplitude and one reproduces the result 
$\widetilde W_C^{(IX)}(r) = (g_A^6/2048\pi^3 f_\pi^6 r^7)\,
e^{-3x}(1+x)(4+5x+3x^2)$ of 
ref.\cite{dreip}. For the isovector spin-spin and tensor potentials generated
by class\,IX one repeats (in a slightly modified form) the technique used
previously for class\,VIII and one ends up with the following representations, 
\begin{eqnarray} \widetilde W_S^{(IX)}(r)&=& {g_A^6 \over 3(16\pi f_\pi^2)^3
r^7} \bigg\{ e^{-3x}(2+2x+x^2)(2-2x-4x^2-x^3)- -\!\!\!\!\!\! 
\int_{-\infty}^{+\infty} {d\zeta_1d\zeta_2 \over \pi^2 \zeta_1\zeta_2 } \,
\nonumber \\ && \times {\partial^2 \over \partial \zeta_2^2} \Big[e^{-x \Sigma}
\Big(12x^{-2}+12x^{-1}\Sigma+2\Sigma^2+8Q+12\beta^2+x((2\Sigma+12\beta)Q-6\Pi) 
\nonumber \\ && +x^2(2\Sigma \Pi+4\beta^2(\alpha^2+3\alpha\gamma+ \gamma^2)) 
+4x^3(Q-\alpha \gamma)\Pi +x^4\Pi^2 \Big)\Big]\bigg\}\,,\end{eqnarray} 
\begin{eqnarray} \widetilde W_T^{(IX)}(r)&=& {g_A^6 \over 3(16\pi f_\pi^2)^3
r^7} \bigg\{e^{-3x}(2+2x+x^2)(2+x-x^2-x^3) \nonumber \\ &&  - -\!\!\!\!\!\! 
\int_{-\infty}^{+\infty}{d\zeta_1d\zeta_2\over \pi^2 \zeta_1\zeta_2} \,
{\partial^2 \over \partial \zeta_2^2} \Big[e^{-x \Sigma} \Big(2\Sigma^2-4Q
+x(2\Sigma Q-6\Pi) \nonumber \\ &&  +x^2 (Q^2+2\alpha^2\gamma^2)+x^3
(Q+2\alpha\gamma)\Pi+x^4 \Pi^2 \Big) \Big] \bigg\}\,. \end{eqnarray} 
Again, an analogous (double-integral) representation for the isoscalar central
potential $\widetilde V_C^{(IX)}(r)$ turned out to be in perfect numerical 
agreement with  the result of ref.\cite{dreip} based on the spectral-function
method. 

\bigskip
\bigskip

\bild{figu9.epsi}{16}
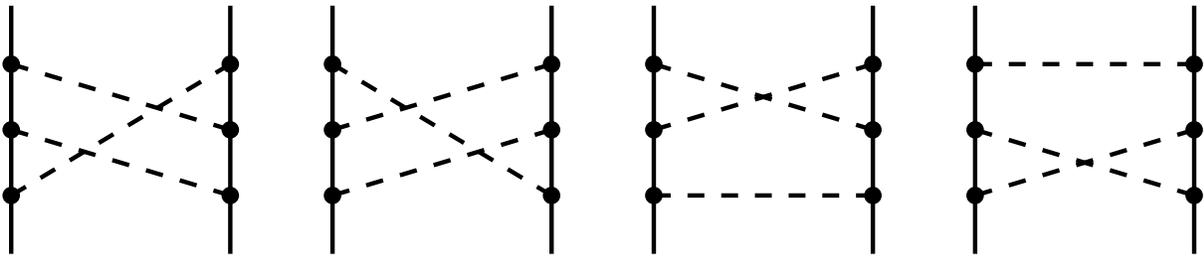
{\it Fig.7: $3\pi$-exchange diagrams of class\,IX proportional to $g_A^6$.} 

\bigskip

Numerical values for the spin-spin and tensor NN-potentials generated by the
diagrams of classes\,VIII and IX are presented in Table\,1. With a typical 
strength of a few MeV (or less) at $r=1.0$\,fm they are comparable to the other
leading order chiral $3\pi$-exchange potentials calculated in ref.\cite{dreip}.

In summary, we have completed in this work the calculation of the leading order
chiral $3\pi$-exchange NN-potentials. Furthermore, we investigated the dominant
next-to-leading order contributions. The (analytical) results presented here
and in refs.\cite{3pipot,dreip} are parameterfree and in a form such that they 
could be easily implemented in a future empirical analysis of low-energy 
elastic NN-scattering. This way one would be able to determine the maximal 
range of validity $r\geq r_0$ of chiral multi-pion exchange in the
NN-interaction.  

\begin{table}[hbt]
\begin{center}
\begin{tabular}{|c|ccccccc|}
    \hline

    $r$~[fm]&0.8&0.9&1.0&1.1&1.2&1.3 &1.4\\
   \hline $\widetilde W_S^{(X)}$~[MeV]
& 3.27 &1.20&0.48&0.21& 0.096&0.046&0.023\\ 
 $\widetilde W_T^{(X)}$~[MeV]
& -20.23 &-7.33&-2.92&-1.26&-0.58&-0.28& -0.14 \\ \hline   
 $\widetilde V_S^{(XII)}$~[MeV]
& 36.30 &13.23&5.30&2.29& 1.05&0.51&0.26\\ 
 $\widetilde V_T^{(XII)}$~[MeV]
& -18.15 &-6.62&-2.65&-1.14&-0.53&-0.25& -0.13 \\  
 $\widetilde W_S^{(XII)}$~[MeV]
& 17.94 &6.69&2.74&1.21& 0.57&0.28&0.15\\ 
 $\widetilde W_T^{(XII)}$~[MeV]
& 27.02 &10.00&4.07&1.78&0.83&0.41& 0.21 \\ \hline   
$\widetilde V_C^{(XIII)}$~[MeV]
& -250.0 &-90.75&-36.21&-15.59& -7.14&-3.45&-1.74\\ 
 $\widetilde V_S^{(XIII)}$~[MeV]
& -112.7 &-41.00&-16.39&-7.07&-3.25&-1.57  & -0.79 \\    
$\widetilde V_T^{(XIII)}$~[MeV]
& 136.4 &49.89&20.06&8.70& 4.02&1.96&0.99\\ 
 $\widetilde W_C^{(XIII)}$~[MeV]
& 333.3 &121.0&48.28&20.78&9.52&4.59& 2.32 \\ 
 $\widetilde W_S^{(XIII)}$~[MeV]
& 54.20 &19.58&7.77&3.33& 1.52&0.73&0.36\\ 
 $\widetilde W_T^{(XIII)}$~[MeV]
& -442.5 &-162.9&-65.92&-28.79&-13.38&-6.55& -3.35 \\ \hline    
$\widetilde V_C^{(XIV)}$~[MeV]
& 250.0 &90.75&36.21&15.59& 7.14&3.45&1.74\\ 
 $\widetilde V_S^{(XIV)}$~[MeV]
& 61.36 &22.80&9.33&4.12&1.94&0.96& 0.50 \\ 
$\widetilde V_T^{(XIV)}$~[MeV]
& 3.13 &1.40&0.66&0.33& 0.17&0.089&0.049\\ 
 $\widetilde W_S^{(XIV)}$~[MeV]
& -8.78 &-3.17&-1.26&-0.54&-0.24&-0.12& -0.058 \\  
$\widetilde W_T^{(XIV)}$~[MeV]
& 10.27 &3.78&1.53&0.67& 0.31&0.15&0.077\\ 
   \hline 
 $\widetilde W_S^{(VIII)}$~[MeV]
& -6.01 &-2.43 &-1.07&-0.502 &-0.248&-0.129 & -0.069 \\  
$\widetilde W_T^{(VIII)}$~[MeV]
& 15.4 & 6.32 &2.82& 1.35 & 0.678&0.357 &0.196\\ 
   \hline
$\widetilde V_S^{(IX)}$~[MeV]
& -2.03 &-1.54 & -0.987&-0.607 &-0.371 &-0.229 &-0.143  \\  
$\widetilde V_T^{(IX)}$~[MeV]
& 7.72 &2.80 &1.09 &0.443 &0.186 &0.079 &0.033\\ 
$\widetilde W_S^{(IX)}$~[MeV]
& 3.52 &1.37 & 0.584& 0.267 &0.129 &0.066 &0.035  \\  
$\widetilde W_T^{(IX)}$~[MeV]
& -0.482 &-0.188 &-0.079 &-0.035 &-0.017 &-0.008 &-0.004\\ 
  \hline
  \end{tabular}
\end{center}
{\it Tab.1: Numerical values of the local NN-potentials generated by the
chiral $3\pi$-exchange graphs of classes\,X,\,XII,\,XIII,\,XIV,\,VIII,\,IX 
(shown in Figs.\,1,\,3,\,4,\,5,\,6,\,7) versus the nucleon distance $r$. The
units of these potentials are MeV.}
\end{table}

\end{document}